%% file: main.tex
\RequirePackage[mathlines]{lineno} % Display line numbers
\documentclass[a4paper,11pt]{article}
\usepackage{jheppub} % for details on the use of the package, please
\usepackage[T1]{fontenc} % if needed
\usepackage{booktabs}
\usepackage{threeparttable}%This package is used for tablenotes
\usepackage{multirow}
\usepackage{subfigure}
\usepackage{float}
\usepackage{textcomp} %just for degree
\usepackage{tikz-feynman}
\usepackage{hyperref}%reference

\let\oldequation\equation
\let\oldendequation\endequation
\renewenvironment{equation}
  {\linenomathNonumbers\oldequation}
  {\oldendequation\endlinenomath}

\def \jpsi {J/\psi}
\def \dsenu{D_{s}^{-}e^{+}\nu_e}
\def \to   {\rightarrow}
\def \sig  {\jpsi\to\dsenu}

\def \ks   {K_S^0}
\def \kp   {K^+}
\def \km   {K^-}

\def \piz  {\pi^0}
\def \pip  {\pi^+}
\def \pim  {\pi^-}

\def \gev  {~\mbox{GeV}}

\def \mevcc{~\mbox{MeV/$c^2$}}

\def \ds   {D_s}

\def \dsm  {D_s^-}
\def \psip {\psi(3770)}
\def \dz   {D^0}
\def \dzb  {\bar{D}^0}
\def \siga {\ks\km}
\def \sigb {\kp\km\pim}
\def \sigc {\kp\km\pim\piz}
\def \sigd {\ks\km\pip\pim}

\def \mds  {M(\ds^-)}
\def \dschiq  {\chi_{1\rm C}^{2}}
\def \pepmiss {|\vec{P}_{e^+}|+|\vec{P}_{\rm{miss}}|}
\def \eop     {E_{e^+}/|\vec{P}_{e^+}|}
\def \egam    {E_{\gamma}^{\rm{tot}}}
\def \pmiss   {\vec{P}_{\rm{miss}}}

\def \umiss    {U_{\rm miss}}

\newcommand{\BESIIIorcid}[1]{\href{https://orcid.org/#1}{\hspace*{0.1em}\raisebox{-0.45ex}{\includegraphics[width=1em]{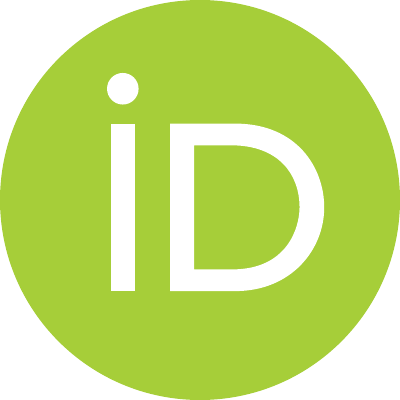}}}}

\begin{document}
% \setrunninglinenumbers
% \begin{linenumbers}

\title{\boldmath Search for the charmonium semi-leptonic weak decay $J/\psi\rightarrow D_s^-e^+\nu_e+c.c.$}

\collaborationImg{\includegraphics[height=30mm,angle=90]{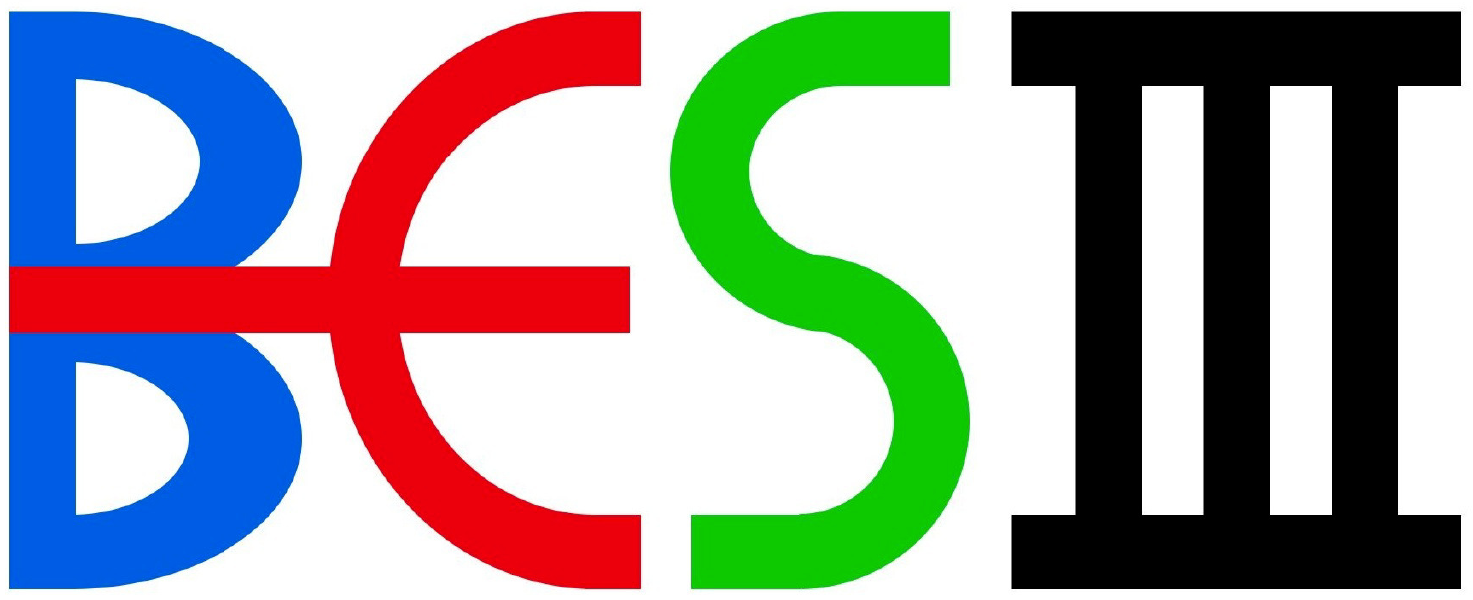}}
\collaboration{The BESIII collaboration}
\emailAdd{besiii-publications@ihep.ac.cn}

\abstract{
Using a data sample of $(10087 \pm 44) \times 10^6$ $J/\psi$ events collected with the BESIII detector at a centre-of-mass energy of $\sqrt{s}=3.097\ \textrm{GeV}$, a dedicated search for the charmonium semileptonic weak decay $J/\psi\rightarrow D_s^-e^+\nu_e + \text{c.c.}$ is performed. No significant signal is observed. An upper limit on the branching fraction is set at $\mathcal{B}(J/\psi\rightarrow D_s^- e^+ \nu_e + \text{c.c.}) < 9.9 \times 10^{-8}$ at the 90\% confidence level. This result improves upon previous constraints by an order of magnitude, representing the most stringent experimental limit to date. It thus provides a critical test of Standard Model predictions and new physics scenarios in heavy-quark dynamics.
}


\arxivnumber{2510.25100}

\maketitle
\flushbottom

\section{Introduction}
\label{sec:introduction}
\hspace{1.5em}
Decays of the $\jpsi$ are predominantly governed by strong or electromagnetic interactions, making its weak decays exceptionally rare.
Moreover, the mass of the $\jpsi$ lies below the open-charm threshold, restricting its weak decays to a single charm meson.
The importance of studying semi-leptonic weak decays of the $\jpsi$ as probes of heavy-quark dynamics was already highlighted long ago~\cite{Sanchis-Lozano:1993vyw}.
In the Standard Model (SM), the inclusive branching fraction (BF) for semi-leptonic $\jpsi$ decays is expected to be of order $\mathcal{O}(10^{-8})$.
However, experimental investigation of these decays remains challenging.
To date, significant efforts have been devoted to heavy-flavour systems such as $D$/$B$ mesons and $\Lambda_c$/$\Xi_c$ baryons, while semi-leptonic decays of the $\jpsi$ have received relatively little attention.

\begin{figure}[htbp] \centering
	\setlength{\abovecaptionskip}{-1pt}
	\setlength{\belowcaptionskip}{10pt}
	\includegraphics[width=0.5\linewidth]{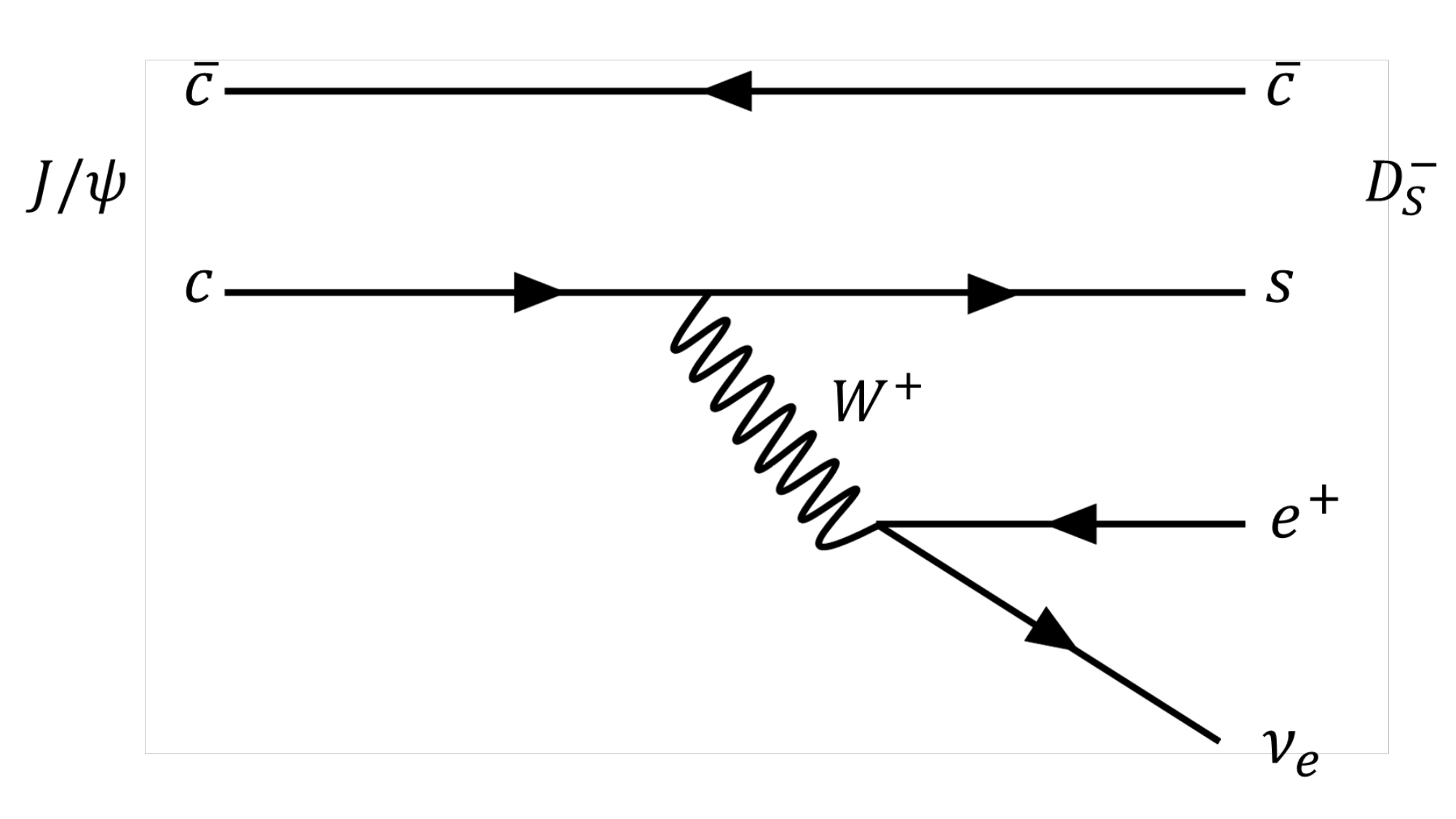}
	\caption{Feynman diagram for $\sig$ at tree level.}
	\label{fig:feynman}
\end{figure}

Among all semi-leptonic $\jpsi$ decays, $\sig$ (charge-conjugate modes are implied throughout) stands out due to its experimental and theoretical advantages.
Figure~\ref{fig:feynman} illustrates a representative tree-level Feynman diagram for this process.
As a Cabibbo-favoured transition, this decay exhibits a significantly larger BF than other channels~\cite{pdg:2022}.
In addition, the final-state positron suffers from substantially less background than its muon counterpart.
% %%%%%%%%%%%%%%%%%%%%%%%%
% \vspace{-0.0cm}
% \vspace{-0.0cm}
% %%%%%%%%%%%%%%%%%%%%%%%%%

Current theoretical predictions for $\mathcal{B}(\sig)$ are based on a variety of approaches, including
Lattice QCD (LQCD)~\cite{meng:2024}, QCD sum rules (QCDSR)~\cite{wang:2008b},
covariant light-front quark model (CLFQM)~\cite{shen:2008, Sun:2023uyn}, Bauer-Stech-Wirbel (BSW) model~\cite{dhir:2013}, covariant constituent quark model (CCQM)~\cite{ivanov:2015},
and Bethe-Salpeter (BS) method~\cite{tian:2017}.
These predictions, summarised in Table~\ref{tab:prediction}, yield BFs up to $\mathcal{O}(10^{-9})$.
The Beijing Spectrometer III~(BESIII) detector has collected huge number of $\jpsi$ events,
offering an unprecedented opportunity to search for this decay.
Notably, certain beyond-SM scenarios-such as top-colour models, $R$-parity-violating minimal supersymmetric extensions, and two-Higgs-doublet models-could enhance the inclusive semi-leptonic $\jpsi$ BF to $\mathcal{O}(10^{-5})$~\cite{hill:1995, datta:1999, zhang:2001, li:2012}.
A search for $\sig$ using the full BESIII dataset therefore provides an important test of SM predictions and can set stringent constraints on new-physics parameters.
%%%%%%%%%%%%%%%%%%%%%%%%%%%%%%%%%%%%%%%%%
\begin{table}[htp]
\caption{Theoretical predictions on the branching fraction of $\sig$.}
\setlength{\tabcolsep}{22pt}
\centering
\begin{tabular}{cc}
\hline \hline
\specialrule{0em}{1pt}{1pt}
    Theoretical model & $\mathcal{B}$~($10^{-11}$)\\
    \hline
    \specialrule{0em}{1pt}{1pt}
    LQCD method~\cite{meng:2024} & $19.0\pm0.8$\\
    \specialrule{0em}{1pt}{1pt}
    QCDSR~\cite{wang:2008b} & $18_{-5}^{+7}$\\
    \specialrule{0em}{1pt}{1pt}
    CLFQM~(2008)~\cite{shen:2008} & $53\sim58$\\
    \specialrule{0em}{1pt}{1pt}
    CLFQM~(2024)~\cite{Sun:2023uyn} & $102.1^{+0.2+0.5}_{-0.2-1.5}$\\
    \specialrule{0em}{1pt}{1pt}
    BSW~\cite{dhir:2013} & $104.0_{-7.5}^{+9.0}$\\
    \specialrule{0em}{1pt}{1pt}
    CCQM~\cite{ivanov:2015} & $33$\\
    \specialrule{0em}{1pt}{1pt}
    BS method~\cite{tian:2017} & $36.7_{-4.4}^{+5.2}$\\
    \specialrule{0em}{1pt}{1pt}
\hline \hline
\end{tabular}
\label{tab:prediction}
\end{table}
%%%%%%%%%%%%%%%%%%%%%%%%%%%%%

Previous experimental searches at BES and BESIII have established limits on several $\jpsi$ weak decays~\cite{bes:2006, bes:2008, bes3:2014_Yuan, bes3:2014, bes3:2017, bes3:2021, bes3:2023, bes3:2023_Wang, bes3:FCNC}.
For $\sig$, the current upper limit~(UL) on the BF is $\mathcal{B} < 1.3 \times 10^{-6}$ at 90\% confidence level~(C.L.), based on $225.3 \times 10^6$ $\jpsi$ events collected with BESIII~\cite{bes3:2014}.

In this work, we search for $\sig$ using $(10087 \pm 44) \times 10^6$ $\jpsi$ events~\cite{bes3:totJpsiNumber, bes3:dataset} collected at $\sqrt{s} = 3.097\gev$ with the BESIII detector.
The $\ds^-$ meson is reconstructed through four hadronic decay modes: $\ds^- \to \siga$, $\ds^- \to \sigb$, $\ds^- \to \sigc$, and $\ds^- \to \sigd$.
To avoid bias in the analysis, we adopted a semi-blind analysis strategy, using approximately 10\% of the $\jpsi$ data sample to validate all analysis procedures before performing the final analysis on the full dataset. 
The validation dataset is not used in deciding selection criteria or in the extraction of PDF shapes and final results.

\section{BESIII detector and Monte Carlo simulation}
\label{sec:detector}
\hspace{1.5em}
The BESIII detector~\cite{Ablikim:2009aa} records symmetric $e^+e^-$ collisions provided by the BEPCII storage ring~\cite{Yu:IPAC2016-TUYA01} in the center-of-mass energy range from 1.84 to 4.95~GeV, with a peak luminosity of $1.1 \times 10^{33}\;\text{cm}^{-2}\text{s}^{-1}$ achieved at $\sqrt{s} = 3.773\;\text{GeV}$.
BESIII has collected large data samples in this energy region~\cite{Ablikim:2019hff}.
The cylindrical core of the BESIII detector covers 93\% of the full solid angle and consists of a helium-based
 multilayer drift chamber~(MDC), a time-of-flight
system~(TOF), and a CsI(Tl) electromagnetic calorimeter~(EMC),
which are all enclosed in a superconducting solenoidal magnet
providing a 1.0~T magnetic field.
The magnetic field was 0.9~T in 2012, which affects 10.79\% of the total $J/\psi$ data.
The solenoid is supported by an octagonal flux-return yoke with resistive plate counter muon identification modules interleaved with steel.
The charged-particle momentum resolution at $1~{\rm GeV}/c$ is
$0.5\%$, and the ${\rm d}E/{\rm d}x$ resolution is $6\%$ for electrons from Bhabha scattering.
The EMC measures photon energies with a resolution of $2.5\%$ ($5\%$) at $1$~GeV in the barrel (end cap) region.
The time resolution in the plastic scintillator TOF barrel region is 68~ps, while that in the end cap region was 110~ps.
The end cap TOF system was upgraded in 2015 using multigap resistive plate chamber technology, providing a time resolution of 60~ps~\cite{etof0,etof1,etof2}, which benefits 87\% of the data used in this analysis.

Monte Carlo (MC) simulated samples produced with a {\sc
geant4}-based~\cite{geant4} software package, which
includes the geometric description of the BESIII detector~\cite{unity,bes:unity, geo1, geo2,bes:visana,geant3and4} and the
detector response, are used to determine detection efficiencies
and to estimate backgrounds.
The simulation models the beam
energy spread and initial state radiation in the $e^+e^-$
annihilations with the generator {\sc
kkmc}~\cite{kkmc1,kkmc2}.
The inclusive MC sample includes both the production of the $J/\psi$
resonance and the continuum processes incorporated in {\sc
kkmc}~\cite{kkmc1, kkmc2}.
All particle decays are modelled with {\sc
evtgen}~\cite{evtgen,evtgen2} using branching fractions
either taken from the
Particle Data Group~\cite{pdg:2022}, when available,
or otherwise estimated with {\sc lundcharm}~\cite{lundcharm1, lundcharm2}.
Final state radiation
from charged final state particles is incorporated using the {\sc
photos} package~\cite{photos2}.
To estimate the signal detection efficiency, we generated 0.8 million signal MC samples $\sig$ for each reconstruction mode using a DIY model in {\sc evtgen}~\cite{evtgen,evtgen2}.
In the DIY model, the influence of the $\bar{c}$ quark and the spin-flip factor is neglected, and only the effect from the weak interaction $c\to s e^- \nu_e+ c.c.$ process is considered~\cite{bes3:2014, bes3:2021}.

\section{Data analysis}
\label{sec:analysis}
\hspace{1.5em}
The analysis is performed using the BESIII offline software system~\cite{BOSS2006, BOSS2008, BOSS2018}.
The $\ds^-$ meson in the signal decay $\sig$ is reconstructed through four hadronic decay modes:
(a) $\ds^-\to\siga$ with $\ks\to\pip\pim$,
(b) $\ds^-\to\sigb$,
(c) $\ds^-\to\sigc$ with $\piz\to\gamma\gamma$, and
(d) $\ds^-\to\sigd$ with $\ks\to\pip\pim$.
Here, ``mode'' refers to the full decay chain including intermediate decays.
All final-state particles except the $\nu_e$ are detected.
The BF of the signal decay is given by
\begin{equation}
\label{eq:bf}
    \mathcal{B}\left(\sig\right)=\frac{\sum_i^4 (N_i/\epsilon_i)}{\sum_i^4\mathcal{B}_i\times N_{\jpsi}},
\end{equation}
where $N_i$ is the number of signal events for mode~$i$, $\epsilon_i$ is the corresponding signal efficiency,
$\mathcal{B}_i$ is the product BF for the full decay chain of mode~$i$, and $N_{\jpsi}$ is the total number of $\jpsi$ events in the data sample.

Charged tracks detected in the MDC are required to be within a polar angle ($\theta$) range of $|\rm{cos\theta}|<0.93$, where $\theta$ is defined with respect to the $z$-axis,
which is the symmetry axis of the MDC.
For charged tracks not originating from $K_S^0$ decays, the distance of closest approach to the interaction point (IP) must be less than 10\,cm along the $z$-axis~($|V_{z}|$),
and less than 1\,cm in the transverse plane~($|V_{xy}|$).
The total number of charged tracks~(including the signal electron) is required to be 4 for mode (a), (b), and (c), and 6 for mode (d), all requiring net charge zero. 
Particle identification~(PID) for charged tracks combines the measurements of energy deposited in the MDC, and the flight time in the TOF. 
The combined likelihoods~($\mathcal{L}$) under the positron, pion, and kaon hypotheses are obtained. 
The positron candidate is required to satisfy $\mathcal{L}(e) > 0.001$ and $\mathcal{L}(e)/(\mathcal{L}(e)+\mathcal{L}(\pi)+\mathcal{L}(K))>0.8$, and its EMC energy deposit $E_{e^+}$ and the momentum in the MDC $P_{e^+}$ should satisfy $0.92<\eop<1.03$.
The charged kaons and pions not originating from $\ks$ are required to satisfy $\mathcal{L}\left(K\right)>\mathcal{L} \left(\pi\right)$ or $\mathcal{L}\left(\pi\right)>\mathcal{L} \left(K\right)$, respectively.

Each $K_{S}^0$ candidate is reconstructed from two oppositely charged tracks within $|V_{z}|<$ 20~cm.
The two charged tracks are assigned as $\pi^+\pi^-$ without imposing further PID criteria.
They are constrained to originate from a common vertex and are required to have an invariant mass within $|M_{\pi^{+}\pi^{-}} - m_{K_{S}^{0}}|<$ 12~MeV$/c^{2}$, where $m_{K_{S}^{0}}$ is the $K^0_{S}$ nominal mass~\cite{pdg:2022}.
The decay length of the $K^0_S$ candidate is required to be greater than twice of its uncertainty.
% away from the IP.

Photon candidates are identified using isolated showers in the EMC.
The deposited energy of each shower must be more than 25~MeV in the barrel region ($|\cos \theta|< 0.80$) and more than 50~MeV in the end cap region ($0.86 <|\cos \theta|< 0.92$).
To exclude showers that originate from charged tracks, the opening angle subtended by the EMC shower and the position of the closest charged track at the EMC must be greater than 10 degrees as measured from the IP.
To suppress electronic noise and showers unrelated to the event, the difference between the EMC time and the event start time is required to be within [0, 700]\,ns.
The total energy of photons not originating from $\piz$ are requested to be smaller than $0.17\gev$, $0.21\gev$, $0.22\gev$, and $0.14\gev$ for the four $\ds^-$ modes (a, b, c, d), respectively.
Each $\piz$ candidate is reconstructed from two photon candidates with
an invariant mass within $115\mevcc<M_{\gamma\gamma}<150\mevcc$.  A
kinematic fit with $\piz$ mass constraint is performed over all
combinations and the $\chi_{\piz}^2$ should be smaller than 200.
For each event, a one-constraint (1C) kinematic fit is performed by constraining the mass of all the $\ds^-$ candidates to its nominal value, and the chi-square value of the fit is saved as $\chi^2_{\rm 1C}$. 
All possible $\gamma\gamma(\pi^0)$ combinations, $\pi^+\pi^-(\ks)$ candidates and the corresponding $\dsm$ candidates are reconstructed. 
The combination with the minimum $\chi^2_{\rm 1C}$ value is assigned as the reconstructed $\dsm$ and the associated $\gamma\gamma$ or $\pi^+\pi^-$ candidate. 

\begin{figure}
    \centering
    \includegraphics[width=0.47\linewidth]{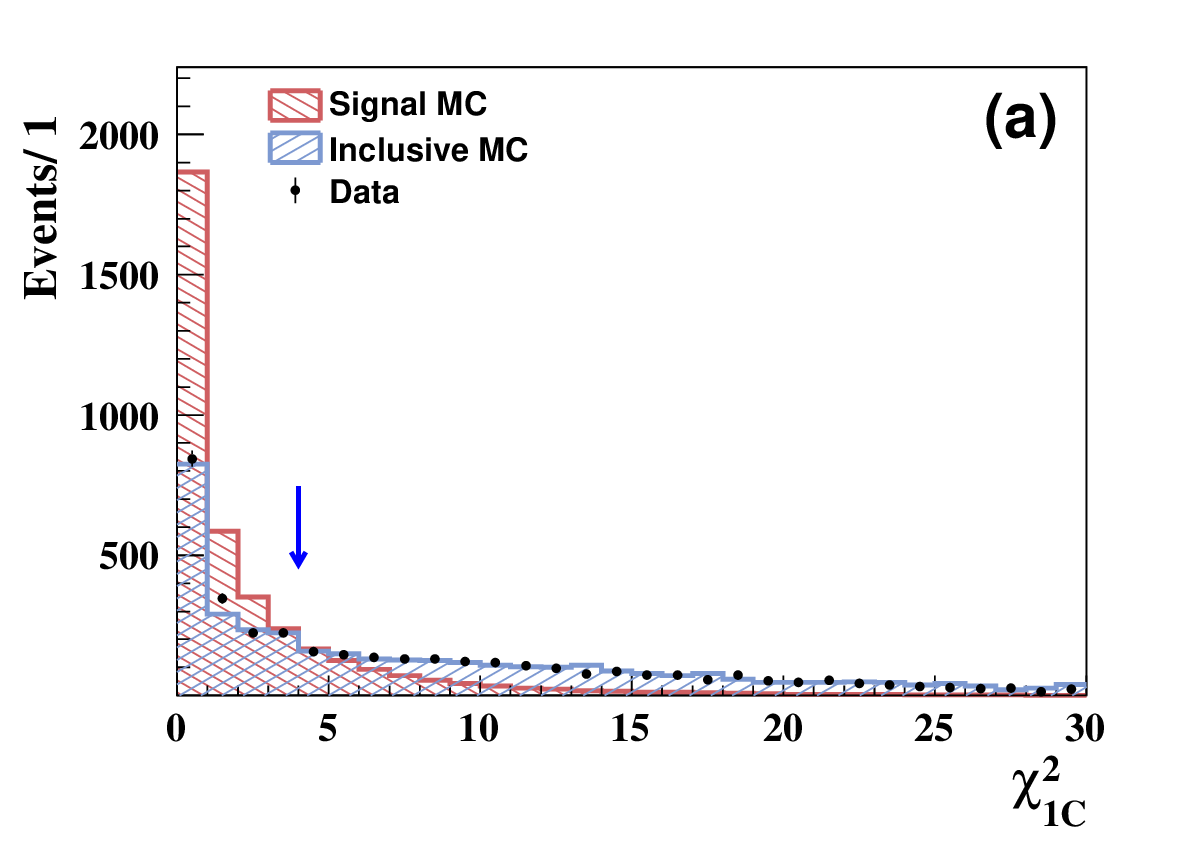}
    \hspace{10pt}
    \includegraphics[width=0.47\linewidth]{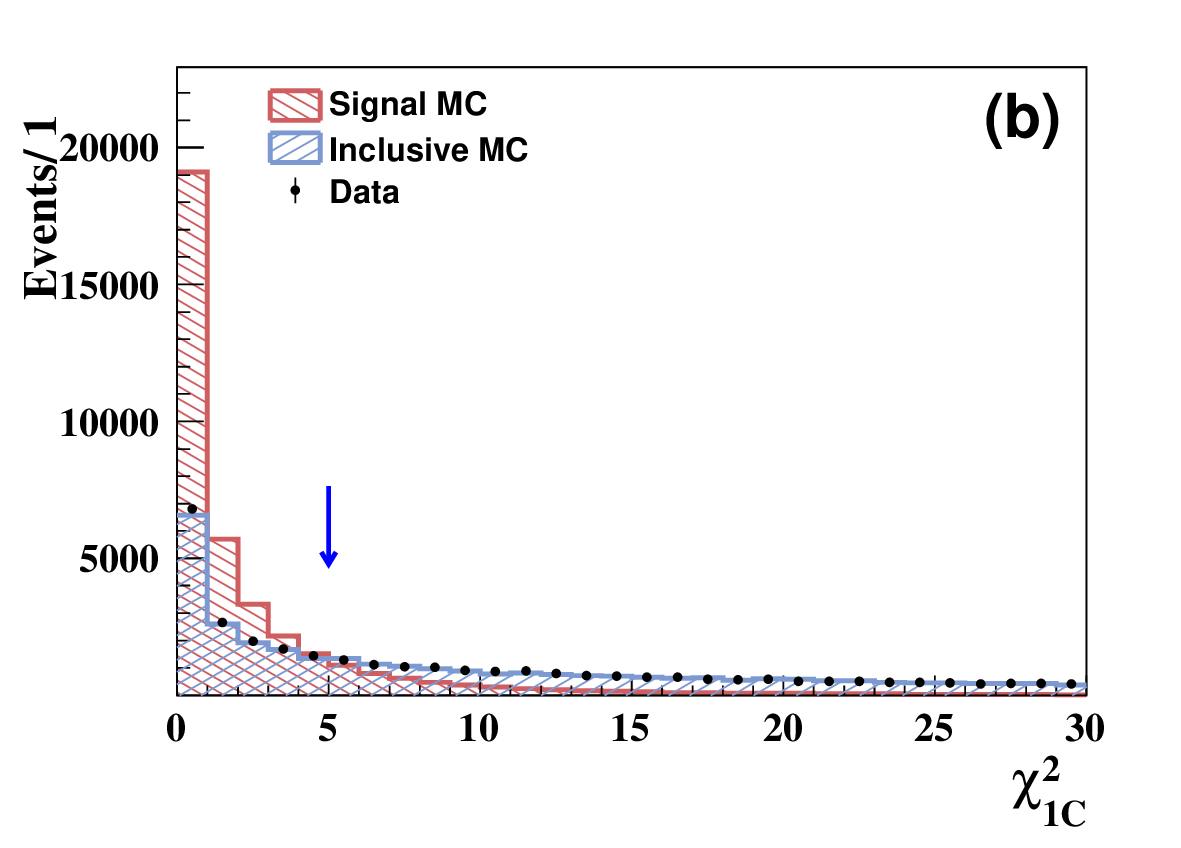}\\
    \includegraphics[width=0.47\linewidth]{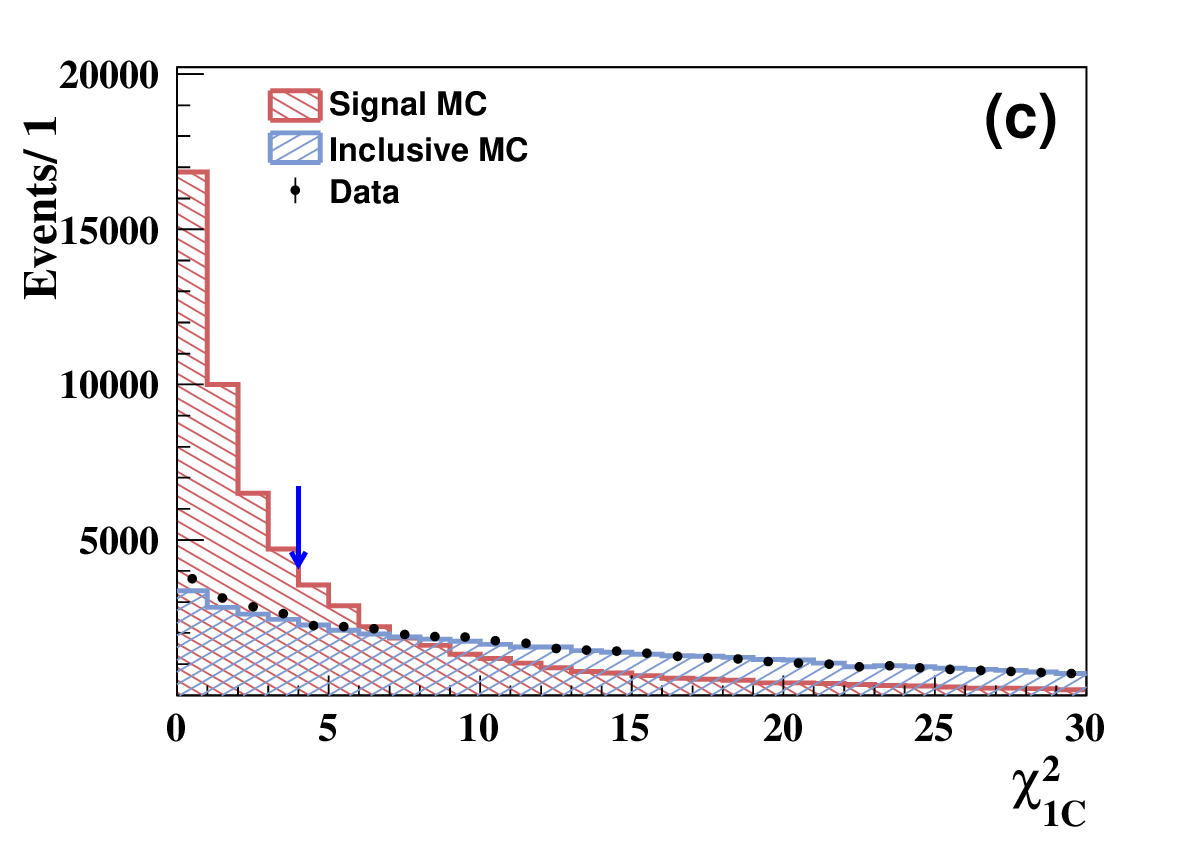}
    \hspace{10pt}
    \includegraphics[width=0.47\linewidth]{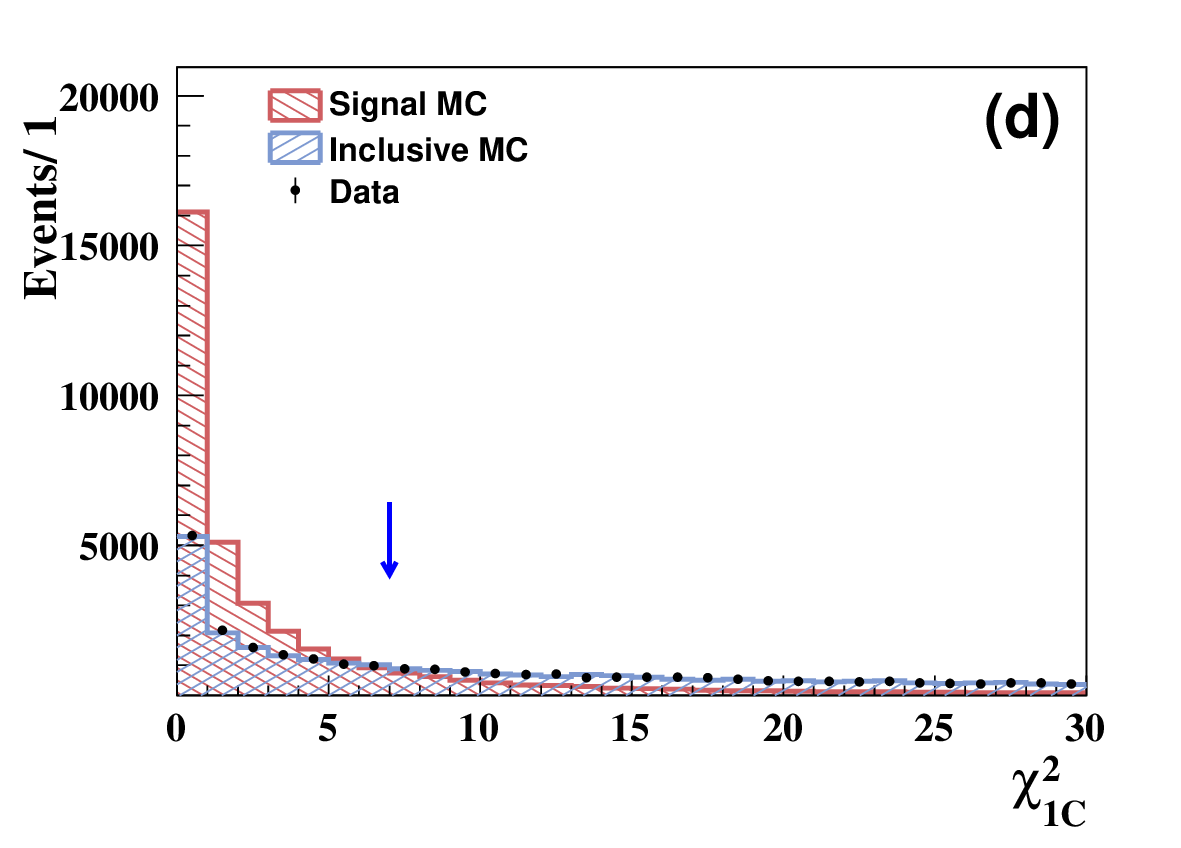}
    \caption{The $\chi_{\rm 1C}^2$ distributions for modes~(a), (b), (c), and (d) are shown in figures~(a), (b), (c), and (d), respectively, with the blue arrows indicating the event selection criteria. In these figures, events satisfy the invariant mass requirement on $\mds$. The signal MC samples are normalised to the number of events in the data.}
    \label{fig:dschisq}
\end{figure}

The invariant mass $\mds$ of the $\ds^-$ candidate is required to be within (1.93, 2.00)~GeV/$c^2$
for mode~(c), and within (1.95, 1.99)~GeV/$c^2$ for the other modes.
% A one-constraint (1C) kinematic fit is performed by constraining the mass of the $\ds^-$ candidate to its known value.
The $\chi^2_{\rm 1C}$ must be less than 4, 5, 4, and 7 for modes~(a)-(d), respectively.
Figure~\ref{fig:dschisq} shows the $\chi^2_{\rm 1C}$ distributions for signal MC, inclusive MC, and data samples after the invariant mass requirement.

In the rest frame of the initial $e^+e^-$ collision, the missing momentum attributed to the undetected
neutrino $\nu_e$ is calculated as $|\vec{P}_{\rm{miss}}| = |\vec{0} - \vec{P}_{\ds^-} - \vec{P}_{e^+}|$,
where $\vec{P}_{\ds^-}$ and $\vec{P}_{e^+}$ are the momenta of the $\ds^-$ meson and the positron, respectively.
To suppress backgrounds with no genuine missing particle, $|\vec{P}_{\rm{miss}}|$ is required to exceed 0.05~GeV/$c$.
Additionally, the sum $|\vec{P}_{e^+}| + |\vec{P}_{\rm miss}|$ must lie between 0.92~GeV/$c$ and 1.15,
1.12, 1.14, or 1.13~GeV/$c$ for modes~(a)-(d), respectively, to enhance kinematic consistency.
% %%%%%%%%%%%%%%%%%%%%%%%%
% \vspace{-0.0cm}

% \vspace{-0.0cm}
% %%%%%%%%%%%%%%%%%%%%%%%%%

Post-selection studies identify two dominant types of background in the inclusive MC sample, each addressed with a dedicated suppression method.
The first background arises from $e^+e^-$ pairs produced in $\piz \to e^+e^-\gamma$ decays or gamma conversions, where the electron is misidentified as a pion from the $\ds^-$ and the positron is incorrectly associated with the $\jpsi$ vertex.
A typical example is $\jpsi \to K^+K^-\piz$ with $\piz \to e^+e^-\gamma$, yielding the final state $K^+K^-e^+e^-\gamma$, which mimics the signal topology $\jpsi \to \ds^-(\sigb)e^+\nu_e$.
This background is suppressed by requiring the opening angle between the positron and any charged pion track to exceed 20 degrees.
This criterion removes nearly all such background events
(reducing the overall background by about 70\%) while retaining approximately 93\% of the signal.
The second background involves a charged pion misidentified as a positron, with an undetected $\piz$ contributing to the missing momentum and mimicking a neutrino.
To reduce this background, mode-dependent invariant mass requirements are applied:
$3.07 < M(K^+K^-\pi^+\pi^-\piz) < 3.12$~GeV/$c^2$ for mode~(b),
$3.07 < M(K^+K^-\pi^+\pi^-\piz\piz) < 3.12$~GeV/$c^2$ for mode~(c), and
$3.08 < M(K^+\pi^+\pi^+\pi^-\pi^-\pi^-\piz) < 3.12$~GeV/$c^2$ for mode~(d).
These requirements remove nearly all of the targeted backgrounds (reducing the overall background by about 50\%) while preserving around 90\% of the signal events.

To derive the information of the missing neutrino, a kinematic variable $\umiss$ is defined as
\begin{equation}
\label{eq:umiss}
    \umiss=E_{\rm miss}-\left|\vec{P}_{\rm miss}\right|c,
\end{equation}
where $E_{\rm miss}$ is the missing energy, calculated as $E_{\rm miss} = 2E_{\rm beam} - E_{\ds^-} - E_{e^+}$ in the $e^+e^-$ centre-of-mass frame.
Here, $E_{\rm beam}$, $E_{\ds^-}$, and $E_{e^+}$ are the energies of the beam, the $\ds^-$ decay products, and the positron, respectively.
Since the neutrino mass is negligible, signal events peak near $\umiss = 0$.
The $\umiss$ distributions for signal MC, inclusive MC, and data are shown in Figure~\ref{fig:fit}.
Some non-uniform background remains in the inclusive MC sample.
The dominant backgrounds are $\jpsi \to K^-\pi^+\pi^+\pi^-\piz$ for mode~(a), $\jpsi \to K^+K^-\pi^+\pi^-\piz$ for mode~(b), $\jpsi \to K^+K^-\pi^+\pi^-\piz\piz$ for mode~(c), and $\jpsi \to K^-\pi^+\pi^+\pi^+\pi^-\pi^-\piz$ for mode~(d).

%%%%%%%%%%%%%%%%%%%%%%%%%%%%%%%%%%%%%%%%%%%%%%%%%%%%%%
\section{Signal extraction}
\label{sec:signal}
\begin{figure}
    \centering
    \includegraphics[width=0.4\linewidth]{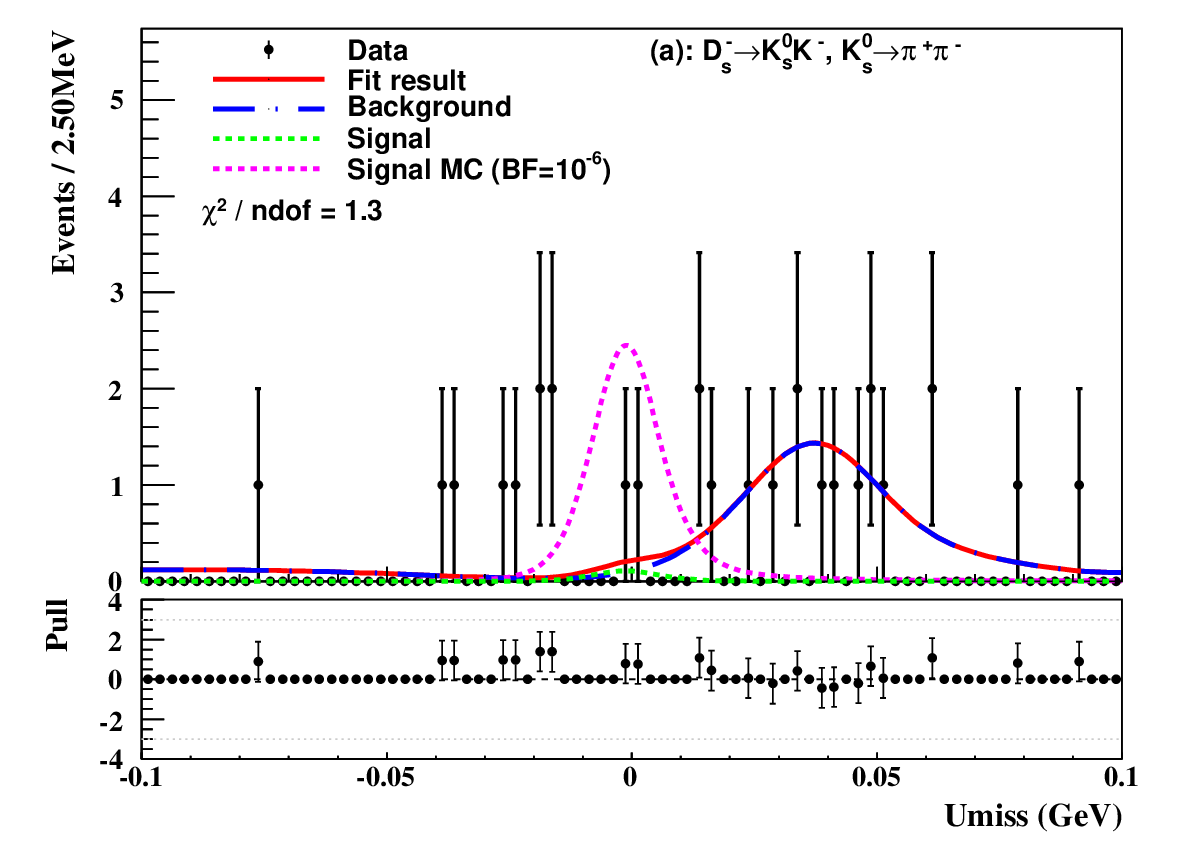}
    \hspace{15pt}
    \includegraphics[width=0.4\linewidth]{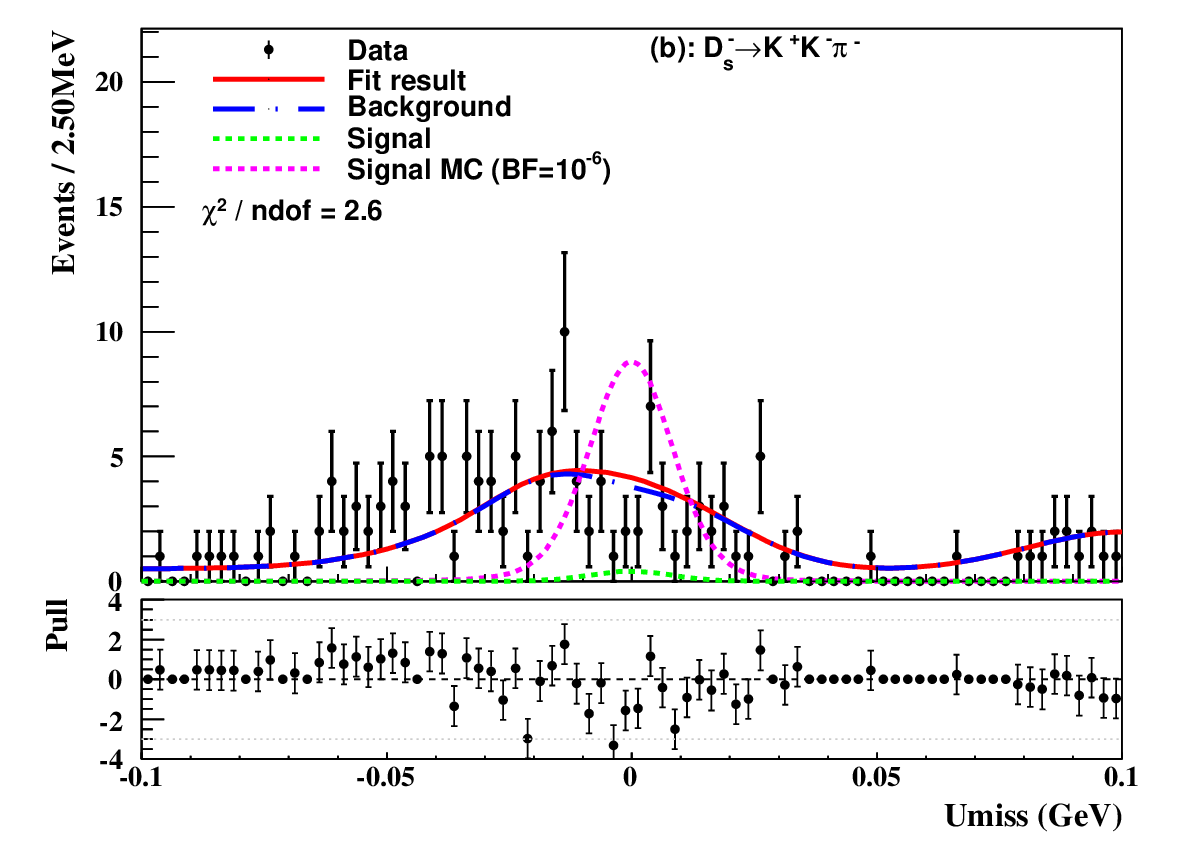}
    \vspace{2pt}
    \\
    \includegraphics[width=0.4\linewidth]{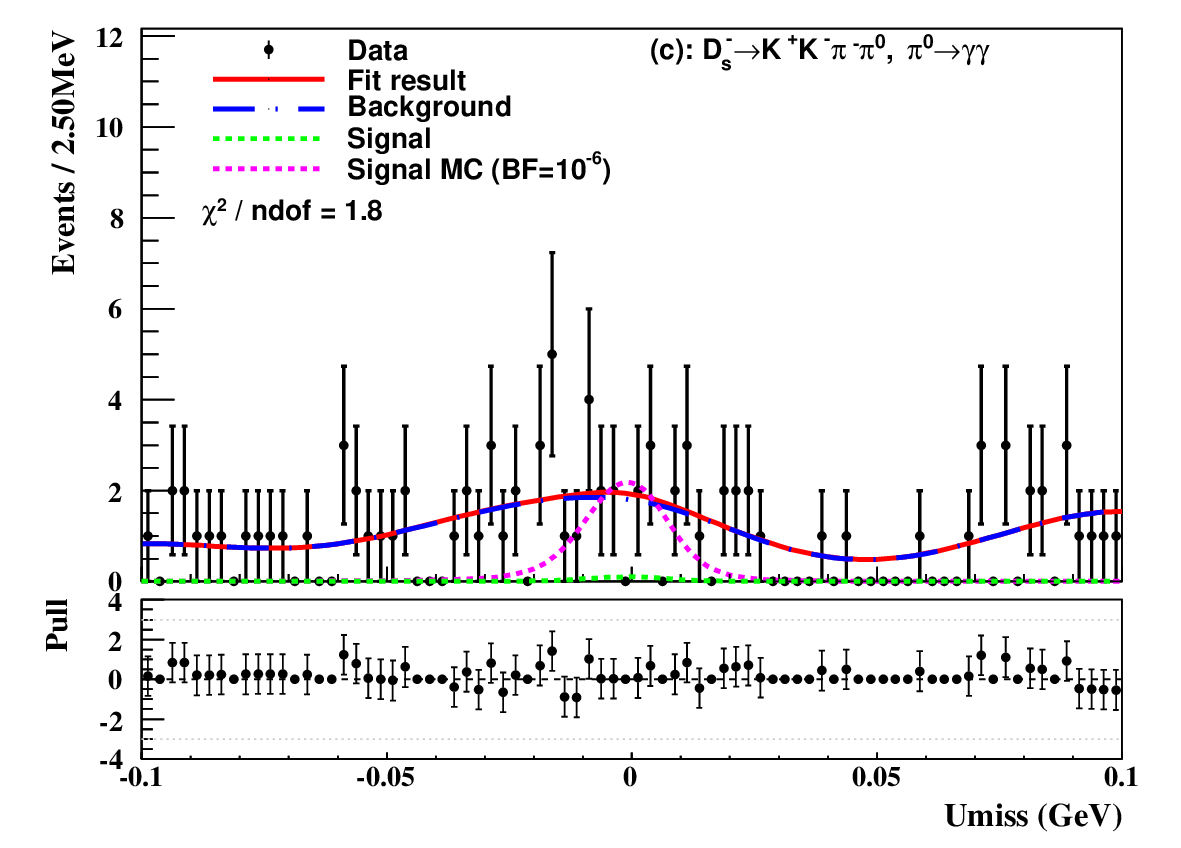}
    \hspace{15pt}
    \includegraphics[width=0.4\linewidth]{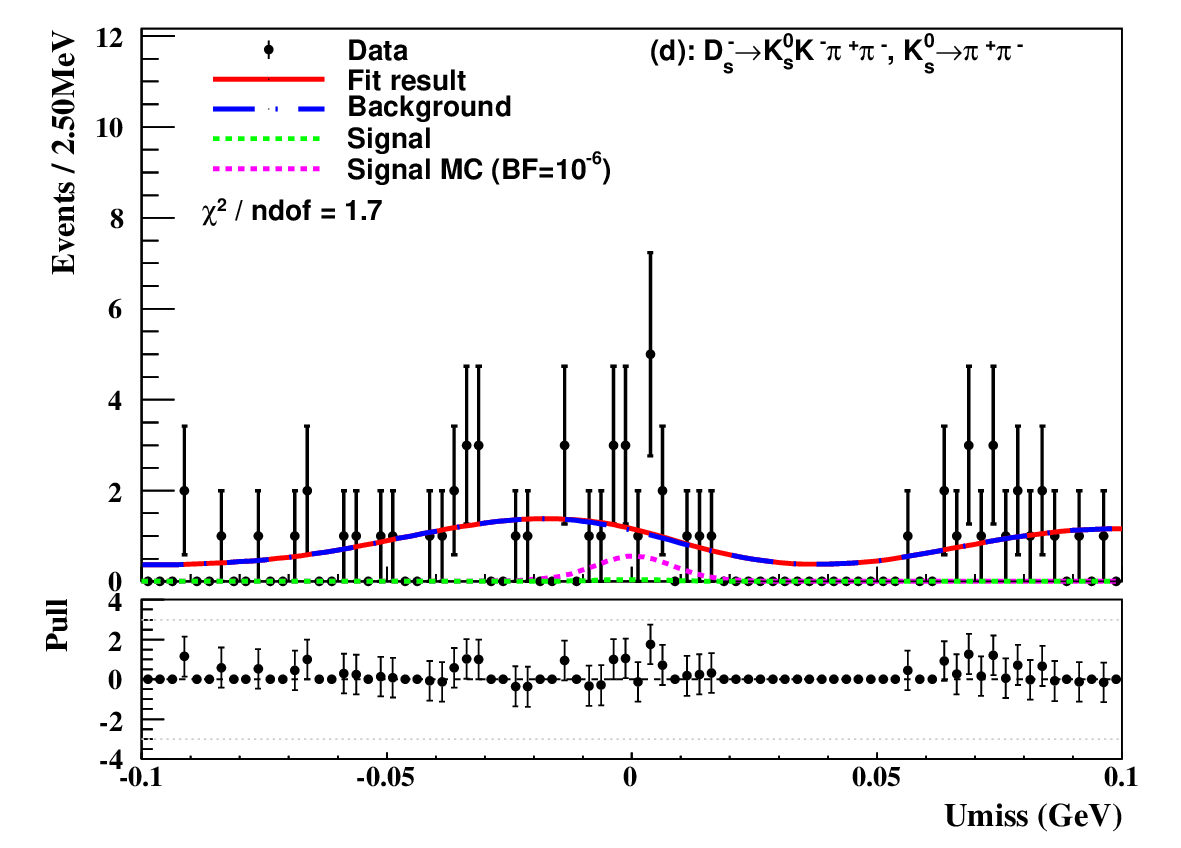}
    \vspace{2pt}
    \\
    \includegraphics[width=0.4\linewidth]{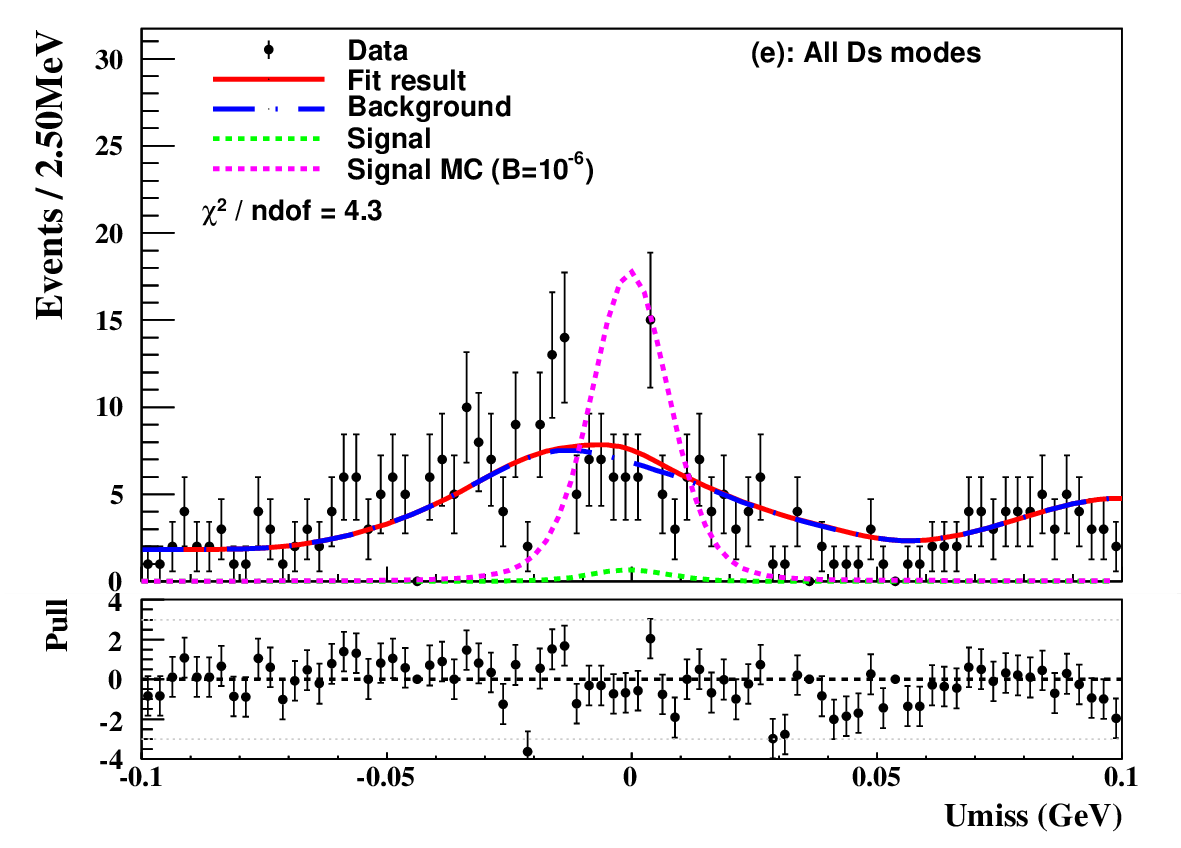}
    \caption{The results of the simultaneous fit to the $\umiss$
      distributions for accepted candidates for modes (a,b,c,d), as well
      as the combined distribution (e) summed over all four modes, are
      presented. The data are shown as black dots with error bars, the
      total fit result as a red curve, the background component as a
      blue dashed line, the extracted signal shape as a green dashed
      line, and the expected signal shape for an assumed branching fraction of
      $\mathcal{B}(\sig)=1\times10^{-6}$ as a pink dashed line. 
      In each panel, a pull distribution is shown below the fit, 
      with the $\chi^2 / \mathrm{ndof}$ value displayed in the fit panel.
    }
    \label{fig:fit}
\end{figure}

\hspace{1.5em}
According to Eq.~\eqref{eq:bf}, the values of $\epsilon_i$ and $\mathcal{B}_i$ are explicitly listed in Table~\ref{tab:sigeff}.
In this table, the signal efficiency is defined as the ratio of events retained after all selection criteria to the total number of events in the signal MC sample.
The signal yields $N_i^{\rm sig}$ for each tag mode are determined through a simultaneous unbinned maximum-likelihood fit to the $\umiss$ distributions for all modes.
In the simultaneous fit, the $N_i^{\rm sig}$ are constrained by
\begin{equation}
\label{eq:simutaneous}
    N_i^{\rm sig}=N_{\rm sig}\times\frac{\mathcal{B}_i\times\epsilon_i}{\sum^4_i\left(\mathcal{B}_i\times\epsilon_i\right)},
\end{equation}
where $N_{\rm sig}$ denotes the total signal yield across all modes.
Similarly, the number of background events for mode~$i$, $N_{i}^{\rm bkg}$, is constrained by
\begin{equation}
\label{eq:simutaneous_incs}
    N^{\rm bkg}_i = N_{\rm bkg}\times \frac{N^{\rm incs}_i}{\sum^4_i N^{\rm incs}_i},
\end{equation}
where $N^{\rm incs}_i$ is the number of inclusive MC events remaining for mode~$i$ after all selections.
The fit function for $\umiss$ distribution of all data events is defined as
\begin{equation}
\label{eq:fit}
    \mathcal{F} = \sum_i^4 N_{i}^{\rm sig}\times\mathcal{PDF}^{\rm sig}_{i}\otimes G\left(\mu_i,\sigma_i\right) + \sum_i^4 N^{\rm bkg}_{i} \times \mathcal{PDF}_{i}^{\rm incs},
\end{equation}
where $\mathcal{PDF}^{\rm sig}_i$ and $\mathcal{PDF}^{\rm incs}_i$ represent the probability density functions derived from the $\umiss$ distributions of the signal MC and inclusive MC samples for mode~$i$, respectively, and $\otimes$ denotes convolution. 
The signal shapes are modeled using a kernel density estimation approach implemented by the ROOT RooKeysPdf class. 
The parameters of the Gaussian function $G(\mu_i,\sigma_i)$ for each mode are fixed to the values given in Table~\ref{tab:congaus}, as obtained from fits to control samples (discussed in detail in Section~\ref{sec:uncertainty}). 
The variation of the Gaussian parameters ($\mu_i, \sigma_i$) among different modes reflects the differences in final-state particle content, particle kinematics, and the associated reconstruction uncertainties. 
These factors result in different detector responses and resolution effects for each mode, and thus require separate correction factors. 
The simultaneous unbinned maximum-likelihood fit to the $\umiss$ distributions for all modes, using the full dataset of $(10087 \pm 44) \times 10^6$ $\jpsi$ events, yields a total signal yield of $N_{\rm sig} = 5.7 \pm 4.5$, where the uncertainty is statistical.
The fit results are shown in Figure~\ref{fig:fit}. 
The validity of this simultaneous fit approach has been verified by performing fits with and without these constraints, 
% as well as to different combinations of all four modes, 
with consistent results found within statistical uncertainties. 

% %%%%%%%%%%%%%%%%%%%%%%%%
% \vspace{-0.0cm}
% %%%%%%%%%%%%%%%%%%%%%%%%%

Since no significant signal is observed, an UL on the signal BF at the
90\% C.L. is set after incorporating all systematic uncertainties in Section~\ref{sec:uncertainty}, using a Bayesian approach~\cite{LB2}.

%%%%%%%%%%%%%%%%%%%%%%%%%%%%%%%%%%%%%%%%%
\begin{table}[htp]
\caption{The branching fractions $\mathcal{B}_i$~\cite{pdg:2022} and signal MC efficiencies $\epsilon_i$ for each mode.}
\label{tab:sigeff}
\setlength{\tabcolsep}{22pt}
\footnotesize
\centering
\begin{tabular}{ccc}
\hline \hline
\specialrule{0em}{1pt}{1pt}
    Mode & $\mathcal{B}_i$ & $\epsilon_i$\\
    \hline
    \specialrule{0em}{1pt}{1pt}
    (a) $\dsm\to\ks(\pip\pim)\km$ & $1.50(1)\%\times69.20(5)\%$ & $18.42\%$ \\
    \specialrule{0em}{1pt}{1pt}
    (b) $\dsm\to\kp\km\pim$ & $5.45(8)\%$ & $14.47\%$ \\
    \specialrule{0em}{1pt}{1pt}
    (c) $\dsm\to\kp\km\pim\piz(\gamma\gamma)$ & $5.53(15)\%\times98.82(3)\%$ & $3.64\%$ \\
    \specialrule{0em}{1pt}{1pt}
    (d) $\dsm\to\ks(\pip\pim)\km\pip\pim$ & $1.57(3)\%\times69.20(5)\%$ & $7.25\%$ \\
    \specialrule{0em}{1pt}{1pt}
\hline \hline
\end{tabular}
\end{table}
%%%%%%%%%%%%%%%%%%%%%%%%%%%%%
%%%%%%%%%%%%%%%%%%%%%%%%%%%%%%%%%%%
\begin{table*}[htp]
\caption{The parameters of the convolved Gaussian function for each mode. They are determined by the control samples of $\psip\to\dz\dzb$. }
\label{tab:congaus}
\centering
\small
\begin{tabular}{ccc}
\hline \hline
        \specialrule{0em}{1pt}{1pt}
        \multirow{2}{*}{Mode} & \multicolumn{2}{c}{Parameter ($\rm MeV$)} \\ \cline{2-3}
        \specialrule{0em}{1pt}{1pt}
                      & Mean value $\mu$ & Sigma value $\sigma$ \\
        \hline
        \specialrule{0em}{1pt}{1pt}
        (a) $\dsm\to\ks\km$ & $0.04\pm0.20$ & $0.35\pm0.43$ \\
        (b) $\dsm\to\kp\km\pim$ & $1.38\pm0.14$ & $4.66\pm0.26$ \\
        (c) $\dsm\to\kp\km\pim\piz$ & $0.72\pm0.09$ & $2.29\pm0.26$ \\
        (d) $\dsm\to\ks\km\pip\pim$ & $2.16\pm0.49$ & $3.56\pm1.08$ \\
        \specialrule{0em}{1pt}{1pt}
\hline \hline
\end{tabular}
\end{table*}
%%%%%%%%%%%%%%%%%%%%%%%%%%%%%%%%%%%%

%%%%%%%%%%%%%%%%%%%%%%%%%%%%%%%%%%%%%%%%%%%%%%%
\section{Systematic uncertainty}
\label{sec:uncertainty}
\hspace{1.5em}
The systematic uncertainties in the determination of the BF comprise two categories: additive uncertainties, which affect the signal yield, and multiplicative uncertainties, which influence the signal efficiency. These two categories of uncertainties propagate to the UL of the BF through different mechanisms.

The additive uncertainties primarily originate from discrepancies between the signal MC and data distributions, and the $\mathcal{PDF}_{\rm sig(incs)}$ extraction, which cannot be directly quantified as precise relative errors and are therefore suppressed through dedicated mitigation strategies.
To suppress the additive uncertainty from signal MC shape, control samples from $\psip\to\dz\dzb$ are utilized, specifically selecting four decay channels that exactly replicate the final state composition of the signal modes.
To account for data-MC differences, we first investigate the differences between control samples of $\psip\to\dz\dzb$ and the corresponding MC samples.
The MC distributions are convolved with a Gaussian whose parameters are determined by fitting to the control samples in data. 
These parameters (summarized in Table~\ref{tab:congaus}) are then applied in the Gaussian convolution of the signal MC, %$\mathcal{PDF}_{\cm{sig}}\otimes G(\mu,\sigma)$,
$\mathcal{PDF}_{\rm{sig}}\otimes G(\mu,\sigma)$,
thereby correcting for observed discrepancies and mitigating the associated systematic uncertainties.
Due to different particle contents and different momentum spectrum, the Gaussian factors are different between different final states. 
% To decrease the effect from the $\mathcal{PDF}_{\rm sig(incs)}$ extraction, multiple extracting parameter are used for comparison, and the method that yields the largest UL result on the BF is selected, ensuring that the most conservative result is used.
To estimate the systematic uncertainty associated with the extraction of the $\mathcal{PDF}_{\rm sig(incs)}$, we compare results obtained using alternative parameter settings in the PDF extracting construction, including variations in the smoothing parameter and the choice of mirroring options for boundary effects. 
The method that yields the largest UL on the BF is adopted to ensure a conservative result.

The multiplicative uncertainties include the relative uncertainties from signal generator model, tracking and PID of charged particles, event selections, and so on. 
These kinds of uncertainties affect the uncertainties of the signal efficiency, which is directly used in the final result calculation~\ref{eq:bf}. 
The detailed values of uncertainties from signal efficiency and other sources for each mode are all shown in Table~\ref{tab:uncertainty1} and Table~\ref{tab:uncertainty2}, respectively.
The uncertainties less than 0.1\% are ignored. 
% The following sources of systematic uncertainty are considered: 
% \begin{itemize}
%     \item Signal generator model: Estimated by comparing the efficiencies from the PHSP and alternative DIY generator models, and the difference is taken as the systematic uncertainty. 
%     \item Tracking efficiency and PID efficiency: This kind of uncertainties have been studied before, considering the correlativity~\cite{pid}. 
%     \item The uncertainties due to MC statistics are negligible. 
%     \item Mass window ($\mds$) and $\dschiq$ requirements: Studied using $\psip\to D^+D^-$ as the control sample, and the efficiency difference between control sample and MC is the assigned uncertainty. 
%     \item Electron identification, gamma reconstruction, and other reconstruction efficiency (for the requirement of 
% \end{itemize}
% A phase space~(PHSP) model is used to estimate the uncertainty accounting for the signal generator model of the $\sig$ decay.
A signal MC sample generated with a theoretical model based on Lattice QCD~\cite{meng:2024} is used to estimate the uncertainty associated with the signal generator model of the $\sig$ decay. 
The systematic uncertainty arising from the theoretical uncertainty of the Lattice QCD-based model is also evaluated, which is about $0.10\%$.
The relative difference in efficiency between the DIY model and the Lattice QCD-based model is assigned as the systematic uncertanty. 
% The relative uncertainty between the efficiencies of the DIY models and PHSP models is assigned as the systematic uncertainty.
The uncertainties due to tracking and PID efficiency for kaons, pions and positrons, as well as the reconstruction of $\ks$ and $\piz$ mesons reconstruction efficiencies have been extensively studied before, considering the correlativity~\cite{pid}.
The uncertainties due to MC statistics are negligible.
To investigate the uncertainty induced by mass window $\mds$ and $\dschiq$ requirements, we use contol samples of $\psip\to D^+D^-$ decays, in which four $D^-$ decays with the similar final states as the four $\dsm$ modes are chosen. 
Systematic uncertainties associated with the $\pepmiss$, $\eop$, $\egam$, and $\pmiss$ requirements are estimated using control samples from $\psip\to\dz\dzb$ decays. 
These uncertainties mainly reflect differences in electron identification efficiency and momentum reconstruction between data and MC.
The chosen $\dz(\dzb)$ decay modes ensure that the final-state particle content closely matches that of the four $\dsm$ modes.
% The data from $\psip\to\dz\dzb$ decays are used as the control samples to estimate the uncertainties caused by the other requirements, including the $\pepmiss$, $\eop$, $\egam$, and $\pmiss$ requirements. 
% This kind of uncertainties are from the electron identification efficiency and momentum reconstruction. 
% Different $\dz(\dzb)$ decays are chosen, which make the total final states of $\psip\to\dz\dzb$ the same as the four $\dsm$ modes. 
The relative uncertainties of four signal modes BFs are considered as the uncertainties~\cite{pdg:2022}.
The relative uncertainty of 0.5\% for $N_{\jpsi}$ is taken as a systematic uncertainty~\cite{totaljpsi}.

The total multiplicative systematic uncertainty is 6.5\%, which will be used to calculate the UL of BF in Section~\ref{sec:result}.

%%%%%%%%%%%%%%%%%%%%%%%%%%%%%%%%%%%
\begin{table*}[tpb]
\caption{The systematic uncertainties (in \%) for each mode. }
\label{tab:uncertainty1}
\centering
\footnotesize
\begin{tabular}{lccccccc}
\hline\hline
\specialrule{0em}{1pt}{1pt}
    Source & Mode (a) & Mode (b) & Mode (c) & Mode (d)\\
    \specialrule{0em}{1pt}{1pt}
    \hline
    \specialrule{0em}{1pt}{1pt}
    MC generator        & $1.3$ & $0.2$ & $0.3$ & $1.5$ \\
    Tracking            & $0.2$ & $1.0$ & $1.5$ & $1.1$ \\
    Particle ID         & $0.2$ & $0.6$ & $0.6$ & $0.6$ \\
    $\ks$ reconstruction & $0.5$ & - & - & $0.8$ \\
    $\piz$ reconstruction & - & - & $1.0$ & - \\
    Signal MC statistics &  - & - &  - & - \\
    $\mds$ requirement    & $0.2$ & $0.8$ & $0.5$ & $0.1$ \\
    $\dschiq$ requirement & - & $2.3$ & $2.7$ & $2.8$ \\
    $\pepmiss$ requirement& $1.0$ & $0.6$ & $0.4$ & $0.2$ \\
    $\eop$ requirement   & $7.0$ & $2.7$ & $3.1$ & $0.1$ \\
    $\egam$ requirement    & $0.3$ & $0.9$ & $0.9$ & $2.2$ \\
    $|\pmiss|$ requirement& - & $0.3$ & - & - \\
    $\theta_{e,\pi}$ requirement & -  & $1.4$ & - & $0.1$ \\
    % Other specific requirement & - & $3.8$ & $2.2$ & $6.3$ \\
    $M(K^+K^-\pi^+\pi^-\piz)$ requirement & - & $3.8$ & - & - \\
    $M(K^+K^-\pi^+\pi^-\piz\piz)$ requirement & - & - & $2.2$ & - \\
    $M(K^+\pi^+\pi^+\pi^-\pi^-\pi^-\piz)$ requirement & - & - & - & $6.3$ \\
    \specialrule{0em}{1pt}{1pt}
    \hline
    \specialrule{0em}{1pt}{1pt}
    $\sigma_{\epsilon_i}$  & $7.2$ & $5.7$ & $5.1$ & $7.5$

% $3.07 < M(K^+K^-\pi^+\pi^-\piz) < 3.12$~GeV/$c^2$ for mode~(b),
% $3.07 < M(K^+K^-\pi^+\pi^-\piz\piz) < 3.12$~GeV/$c^2$ for mode~(c), and
% $3.08 < M(K^+\pi^+\pi^+\pi^-\pi^-\pi^-\piz) < 3.12$~GeV/$c^2$ for mode~(d).
    \\
\hline\hline
\end{tabular}
\end{table*}
%%%%%%%%%%%%%%%%%%%%%%%%%%%%%%%%%%%
%%%%%%%%%%%%%%%%%%%%%%%%%%%%%%%%%%%
\begin{table*}[tpb]
\caption{The multiplicative systematic uncertainties (in \%). }
\label{tab:uncertainty2}
\centering
\footnotesize
\begin{tabular}{lccccccc}
\hline\hline
\specialrule{0em}{1pt}{1pt}
    Source & Mode (a) & Mode (b) & Mode (c) & Mode (d)\\
    \specialrule{0em}{1pt}{1pt}
    \hline
    \specialrule{0em}{1pt}{1pt}
    Signal efficiency $\epsilon_j$ & $7.2$ & $5.7$ & $5.1$ & $7.5$ \\
    BFs of $\ds^-$ tags    & $2.4$ & $1.9$ & $4.4$ & $5.2$ \\
    Total number of $\jpsi$ events & \multicolumn{4}{c}{$0.5$} \\
    \specialrule{0em}{1pt}{1pt}
    \hline
    \specialrule{0em}{1pt}{1pt}
    Total $\sigma_{\rm sys}$ &  \multicolumn{4}{c}{$6.5$} \\
\hline\hline
\end{tabular}
\end{table*}
%%%%%%%%%%%%%%%%%%%%%%%%%%%%%%%%%%%

\section{Result}
\label{sec:result}
\begin{figure}[htbp] \centering
\setlength{\abovecaptionskip}{-1pt}
\setlength{\belowcaptionskip}{10pt}
\includegraphics[width=0.8\linewidth]{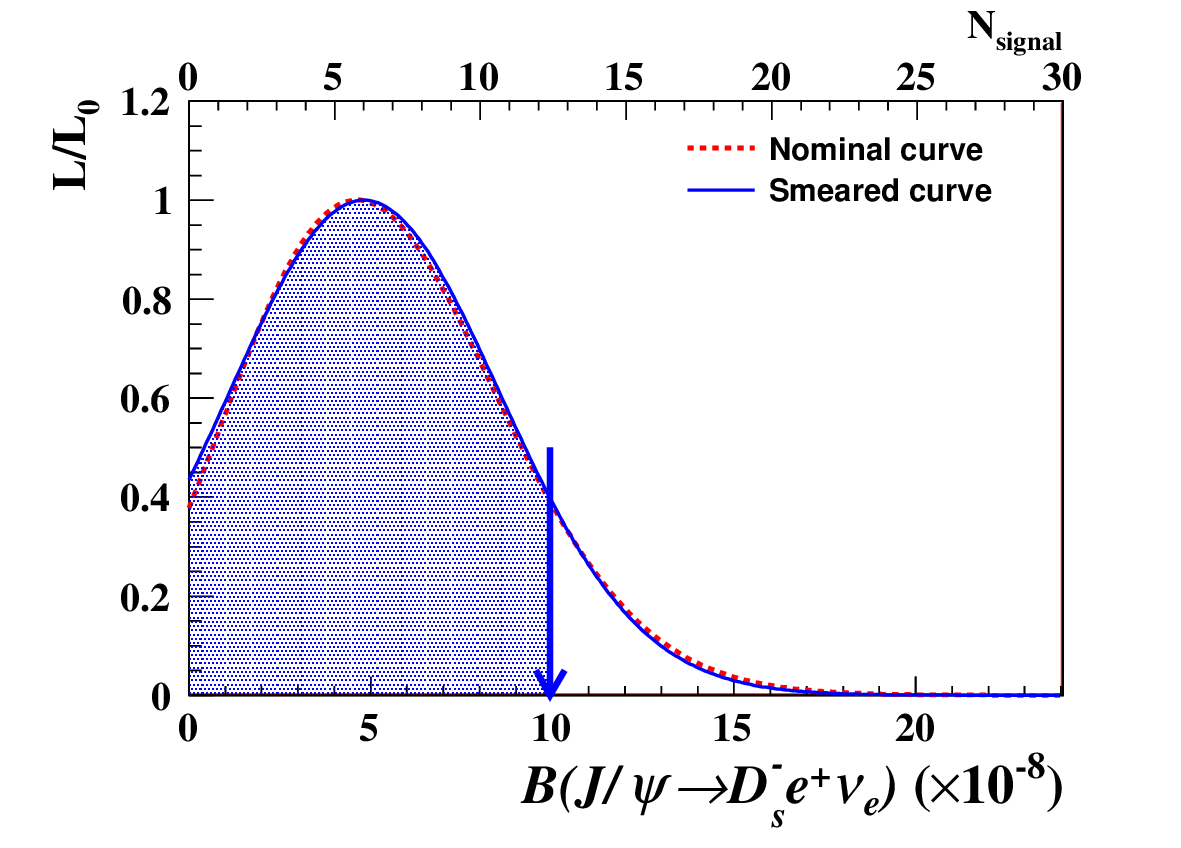}
\caption{The distributions of the normalised likelihoods, both before and after incorporating the systematic uncertainties, are shown. The red dashed line represents the initial likelihood distribution, while the blue line corresponds to the distribution after systematic uncertainties have been taken into account. The blue arrow indicates the upper limit on the branching fraction at the 90\% confidence level, where the integrated area under the blue curve from zero to this point contains 90\% of the total probability above zero.}
\label{fig:result}
\end{figure}

\hspace{1.5em}
To determine the UL, and incorporating the multiplicative uncertainties quantified in Section~\ref{sec:uncertainty}, a Bayesian approach~\cite{LB1, LB2} is employed to compute the likelihood distribution according to Eq.~\eqref{eq:llh}.
In this calculation, the effect of systematic uncertainties is incorporated by convolving the likelihood function with a Gaussian representing the systematic uncertainty.
\begin{equation}
\label{eq:llh}
    \mathcal{L}(\mathcal{B})_{\rm smear} \propto \int^1_0 \exp{\left[-\frac{(\epsilon\mathcal{B}/\hat{\epsilon}-\hat{\mathcal{B}})^2}{2\sigma_{\mathcal{B}}^2}\right]} \times \frac{1}{\sqrt{2\pi}\sigma_\epsilon}\ \exp{\left[-\frac{(\epsilon-\hat{\epsilon})^2}{2\sigma_\epsilon^2}\right]} \rm{d}\epsilon,
\end{equation}
where $\hat{\epsilon}$ is the nominal signal efficiency and $\sigma_\epsilon=\sigma_{\rm sys}\cdot\epsilon$ is the systematic uncertainty of efficiency.
By integrating $\mathcal{L}(\mathcal{B})_{\rm smear}$ from 0 up to 90\% of the physical region ($\mathcal{B}\geq0$) under the smeared curve, we set the UL on the BF to be $\mathcal{B}(\sig) < 9.9\times 10^{-8}$ at the 90\% C.L. using Eq.~\eqref{eq:bf}, as shown in Figure~\ref{fig:result}.

% %%%%%%%%%%%%%%%%%%%%%%%%
% %%%%%%%%%%%%%%%%%%%%%%%%%

\section{Summary}
\label{sec:summary}
\hspace{1.5em}
In summary, a search for the semi-leptonic weak decay $\sig$ is performed using a sample of $(10087 \pm 44) \times 10^6$ $\jpsi$ events collected at $\sqrt{s} = 3.097\gev$ with the BESIII detector. No significant signal is observed.
After incorporating all systematic uncertainties, we set an upper limit on the branching fraction of $\mathcal{B}(\jpsi \to D_s^- e^+\nu_e) < 9.9 \times 10^{-8}$ at the 90\% confidence level.
This result improves upon the previous limit by one order of magnitude~\cite{bes:2006} and remains consistent with Standard Model predictions from various theoretical approaches~\cite{meng:2024,wang:2008b,shen:2008,Sun:2023uyn,dhir:2013,ivanov:2015,tian:2017}, thereby placing more stringent constraints on potential new-physics scenarios in charmonium decays.

\acknowledgments
\hspace{1.5em}
The BESIII Collaboration thanks the staff of BEPCII (https://cstr.cn/31109.02.BEPC) and the IHEP computing center for their strong support. This work is supported in part by National Key R\&D Program of China under Contracts Nos. 2023YFA1606000, 2023YFA1606704; National Natural Science Foundation of China (NSFC) under Contracts Nos. 11635010, 11935015, 11935016, 11935018, 12025502, 12035009, 12035013, 12061131003, 12192260, 12192261, 12192262, 12192263, 12192264, 12192265, 12221005, 12225509, 12235017, 12361141819; the Chinese Academy of Sciences (CAS) Large-Scale Scientific Facility Program; the Strategic Priority Research Program of Chinese Academy of Sciences under Contract No. XDA0480600; CAS under Contract No. YSBR-101; 100 Talents Program of CAS; The Institute of Nuclear and Particle Physics (INPAC) and Shanghai Key Laboratory for Particle Physics and Cosmology; ERC under Contract No. 758462; German Research Foundation DFG under Contract No. FOR5327; Istituto Nazionale di Fisica Nucleare, Italy; Knut and Alice Wallenberg Foundation under Contracts Nos. 2021.0174, 2021.0299; Ministry of Development of Turkey under Contract No. DPT2006K-120470; National Research Foundation of Korea under Contract No. NRF-2022R1A2C1092335; National Science and Technology fund of Mongolia; Polish National Science Centre under Contract No. 2024/53/B/ST2/00975; STFC (United Kingdom); Swedish Research Council under Contract No. 2019.04595; U. S. Department of Energy under Contract No. DE-FG02-05ER41374

\newpage
\input{authorlist_2025-07-18.tex}
% \end{linenumbers}
\end{document}

%% file: authorlist_2025-07-18.tex
%% Saved at => 2025-07-18
M.~Ablikim$^{1}$\BESIIIorcid{0000-0002-3935-619X},
M.~N.~Achasov$^{4,b}$\BESIIIorcid{0000-0002-9400-8622},
P.~Adlarson$^{80}$\BESIIIorcid{0000-0001-6280-3851},
X.~C.~Ai$^{85}$\BESIIIorcid{0000-0003-3856-2415},
R.~Aliberti$^{37}$\BESIIIorcid{0000-0003-3500-4012},
A.~Amoroso$^{79A,79C}$\BESIIIorcid{0000-0002-3095-8610},
Q.~An$^{76,62,\dagger}$,
Y.~Bai$^{60}$\BESIIIorcid{0000-0001-6593-5665},
O.~Bakina$^{38}$\BESIIIorcid{0009-0005-0719-7461},
Y.~Ban$^{48,g}$\BESIIIorcid{0000-0002-1912-0374}
% ...
\\
\vspace{0.2cm}
(BESIII Collaboration)\\
\vspace{0.2cm} {\it
$^{1}$ Institute of High Energy Physics, Beijing 100049, People's Republic of China\\
% ...
$^{4}$ Budker Institute of Nuclear Physics SB RAS (BINP), Novosibirsk 630090, Russia\\
% ...
$^{37}$ Johannes Gutenberg University of Mainz, Johann-Joachim-Becher-Weg 45, D-55099 Mainz, Germany\\
$^{38}$ Joint Institute for Nuclear Research, 141980 Dubna, Moscow region, Russia\\
% ...
$^{48}$ Peking University, Beijing 100871, People's Republic of China\\
% ...
$^{60}$ Southeast University, Nanjing 211100, People's Republic of China\\
% ...
$^{62}$ State Key Laboratory of Particle Detection and Electronics, Beijing 100049, Hefei 230026, People's Republic of China\\
% ...
$^{76}$ University of Science and Technology of China, Hefei 230026, People's Republic of China\\
% ...
$^{79}$ University of Turin and INFN, (A)University of Turin, I-10125, Turin, Italy; (B)University of Eastern Piedmont, I-15121, Alessandria, Italy; (C)INFN, I-10125, Turin, Italy\\
$^{80}$ Uppsala University, Box 516, SE-75120 Uppsala, Sweden\\
% ...
$^{85}$ Zhengzhou University, Zhengzhou 450001, People's Republic of China\\

\vspace{0.2cm}
$^{\dagger}$ Deceased\\
% ...
$^{b}$ Also at the Novosibirsk State University, Novosibirsk, 630090, Russia\\
% ...
$^{g}$ Also at State Key Laboratory of Nuclear Physics and Technology, Peking University, Beijing 100871, People's Republic of China\\
% ...

}
%% ends here %%